\documentclass[pre,twocolumn,amsmath,amsfonts,amssymb,showpacs]{revtex4} %,nofootinbib
\usepackage{graphicx}
\usepackage{mathptmx}      % use Times fonts if available on your TeX system
\usepackage{latexsym}
\def\beq{\begin{equation}}
\def\eeq{\end{equation}}
\def\bea{\begin{eqnarray}}
\def\eea{\end{eqnarray}}

\begin{document}

\title{Condensation of an ideal gas obeying non-Abelian statistic}

\author{Behrouz Mirza}
\email{b.mirza@cc.iut.ac.ir}
\author{Hosein Mohammadzadeh}
\email{h.mohammadzadeh@ph.iut.ac.ir}

\affiliation{Department of Physics, Isfahan University of Technology, Isfahan 84156-83111, Iran}

%%\date{October, 2010}


\begin{abstract}
We consider the thermodynamic geometry of an ideal non-Abelian
gas. We show that, for a certain value of the fractional parameter
and at the relevant maximum value of fugacity, the thermodynamic
curvature has a singular point. This indicates a
condensation such as Bose-Einstein condensation for non-Abelian statistics
 and we work out the phase transition temperature in various dimensions.

\end{abstract}
\maketitle
%%%%%%%%%%%%%%%%%%%%%%%%%%%%%%%%%%%%%%%%%%%%%%%%%%%%%

%%%%%%%%%%%%%%%%%%%%%%%%%%%%%%%%%%%%%%%%%%%%%%%%%%%%%
\section{Introduction}
Bose-Einstein condensation (BEC) is a well-known phenomenon for boson
gas which was first predicted in 1924 \cite{Einestein}  and later
observed in 1995 in a
 remarkable series of experiments on vapors of rubidium and
 sodium \cite{Anderson,Davis}. It has been shown that there is no
 condensation for the nonrelativistic boson particles in the two and one
 spatial dimensions while, for ultra relativistic particles one can
 find a finite phase transition temperature in all
 dimensions $D\geq2$ \cite{May}. A natural
 question to raise here is whether  there are other statistical systems
 that can be condensate and  if they occur
 in low dimensions. It has been shown that an ideal boson
  gas has an upper bound on fugacity, where the thermodynamic
  curvature is singular and where BEC occurs. Recently, we investigated
  the thermodynamic geometry of some systems with intermediate
  statistics and found some evidence for the condensation of nonpure bosonic systems.

   Thermodynamics of different models can be studied by a
   qualitative tool, namely, the
   thermodynamic curvature, which has been introduced by the  theory of thermodynamic
   geometry \cite{Gibbs,Weinhold,Ruppeiner1}. Thermodynamic geometry
    can be used as a measure of
   statistical interaction, e.g., the thermodynamic curvature of  an ideal
   classical gas is zero and it is positive and negative for
   boson and fermion gases, respectively, indicating that
   the statistical interactions of these models are non interacting, attractive, and
   repulsive \cite{Nulton,Mrugala2}. It was argued that the scalar
   curvature could be used to show that fermion gases were more stable than
   boson gases and that, therefore, one can utilize the scalar curvature as a stability
   criterion \cite{Mrugala2}. Also, it has been shown that the
   singular point of thermodynamic curvature  coincides with the
   critical phase transition point of some thermodynamic systems \cite{Mrugala2,brody2}.
   Recently, we worked out the thermodynamic curvature of a thermodynamic
   system in which  particles obey fractional exclusion
   statistics \cite{Mirza1}.
   The study yielded
    some interesting results about the thermodynamic properties of these systems.

Investigation of the thermodynamic geometry of an ideal gas with
particles
 obeying Abelian fractional exclusion statistics has shown that there
 is no phase transition such as Bose-Einstein condensation, while for other
 kinds of intermediate statistics such as Polychronakos and  $q$-deformed
 bosons, the singular point of thermodynamic curvature  coincides  with the
 condensation point \cite{Mirza1}. In this paper, we consider the possibility
 of condensation of an ideal gas with particles obeying non-Abelain exclusion statistics.
 A hint to a possible origin of non-Abelian systems may lie in the observation that
 these systems are fluids in which  microscopic particles
 of a quantum statistics group form clusters whose statistics is effectively bosonic.
 These clusters then form a Bose-Einstein condensate. Generally, breaking of a cluster
 in that condensate costs  energy equal at least to the energy gap. Generally speaking, vortices
 formed in these Bose-Einstein condensates become non-Abelian quasiparticles when their presence
 allows clusters to be broken at no energy cost, leading to ground state degeneracy \cite{stern}.

Haldane introduced the idea of generalized Pauli principle or
fractional exclusion statistics \cite{Haldane}. His definition of
particles obeying the generalized exclusion statistics with
finite dimensional Hilbert space is motivated by such physical
examples as quasiparticles in the fractional quantum Hall system
and spinons in antiferromagnetic spin chains \cite{Laughlin}. The
fractional parameter governing the fractional exclusion
statistics in this case is defined by $g=-\frac{\Delta d}{\Delta
N}$, where $\Delta d$ is the change in the dimension of the
single particle Hilbert space and $\Delta N$ is the change in the
number of particles when the size of the system and the boundary
conditions are unchanged. The fractional parameter is a real
number $0\leq g\leq 1$, and the marginal value $g=0$ $(g=1)$
corresponds to bosons (fermions). It has been shown that spinons
in the Heisenberg chains could also show the non-Abelian
exclusion statistics \cite{Fahm}. It is well known that strongly
correlated many body systems, especially in the low dimensions,
can have quasiparticles that are very different from the
microscopic degrees of freedom from which the system is built.
Also, the braid group in two spatial dimensions may be
represented by the noncommute matrices; the non-Abelian braid
statistics could be  realized in a real world. These include the
quasi particles in the non-Abelian fractional quantum Hall state
as well as those in the conformal field theory
\cite{Moore,Huang0}.

This paper is organized as follows. In Sec. II, we summarized some properties of the non-Abelian statistics
and some thermodynamic quantities of an ideal gas with particles obeying non-Abelian statistics being explored. Thermodynamic geometry and
some interesting properties of this kind of statistics are investigated in Sec.  III. In Sec. IV, we will show
that the singular point of thermodynamic geometry at the upper bound of fugacity coincides with to
the singular point of derivative of particle number and there is a condensation at a related temperature for a
specified value of fractional parameter.
\section{Non-Abelian Statistics}
Using the thermodynamic Bethe ansatz for
specific nondiagonal scattering matrices, the occupation number
distribution function for the particle obeying the non-Abelian
statistics has been derived by Guruswamy and Schoutens
\cite{sathya}. They proposed the following equation for $M$ types
of particles and $k$ types of pseudoparticles:
 \bea
 \left(\frac{\Lambda_{A}-1}{\Lambda_{A}}\right)\prod_{B}\Lambda_{B}^{\alpha_{AB}}\prod_{i}\Lambda_{i}^{G_{Ai}}=z_{A},~~~A=1,\ldots, M,\nonumber\\
 \left(\frac{\Lambda_{i}-1}{\Lambda_{i}}\right)\prod_{A}\Lambda_{A}^{G_{iA}}\prod_{j}\Lambda_{j}^{G_{ij}}=1,~~~~i=1,\ldots, k-1,
 \eea
where $\alpha_{AB}$ is the Abelian part statistics matrix,
$G_{iA}=G_{Ai}=-\frac{1}{2}\delta_{i,1}$ and
$G_{ij}=\frac{1}{2}(C_{k-1})_{ij}$, with $C_{ij}$ as the Cartan
matrix of the associated group. Also, $\Lambda_{a}$ is the
single-level grand canonical partition function and
$z_{a}=\exp[\beta(\mu_{a}-\epsilon)]$  with $\epsilon$ as the
energy
 level and $\mu_{a}$ representing the chemical potential of the particle of type $a$. In the case $G_{iA}=0$, the
  above equations describe the Haldane fractional exclusion statistics. The simplest extension
  of the Abelian to non-Abelian case is related to the $M=1$ and $k=2$, which can describe
  the $q$-pfaffian non-Abelian fractional quantum Hall effect. In this case, the above equations reduce to the following relations \cite{sathya,Huang0}:
 \bea
 \left(\frac{\Lambda-1}{\Lambda}\right)\Lambda^{\alpha}\Lambda_{1}^{-1/2}=z \ ,~~~~~~\left(\frac{\Lambda_{1}-1}{\Lambda_{1}}\right)\Lambda^{-1/2}\Lambda_{1}=1.
 \eea
Eliminating $\Lambda_{1}$ from the above equations yields,
 \bea
 \left(\frac{\Lambda-1}{\Lambda}\right)\Lambda^{\alpha}(1+\Lambda^{1/2})^{-1/2}=z.\label{lambda}
 \eea
Since the distribution function $n$ is defined by $n=\frac{d\ln(\Lambda)}{d\ln(z)}$, we obtain
 \bea
 n(\Lambda)=\frac{1}{\frac{1}{4}\frac{1}{1+\Lambda^{1/2}}+\frac{1}{\Lambda-1}+(\alpha-\frac{1}{4})}.
 \eea

\noindent It is simple to show that the distribution function $n(\Lambda)$ has two different behaviors in the extremal
limits of $z$ which are not exactly similar to the Abelian exclusion statistics.

\bea
 n=\left\{\begin{array}{cc}
 \frac{1}{\alpha-\frac{1}{4}} & \alpha>\frac{1}{4}, ~z\rightarrow\infty \\
 \infty & \ \ \ \  \alpha\le\frac{1}{4},  ~z\rightarrow z_{\max} \\
 \end{array}
 \right.,
\eea

For $\alpha>\frac{1}{4}$,  the maximum population  $\frac{1}{\alpha-\frac{1}{4}}$ is finite and we
do not expect a phase transition. This is similar to the Abelian fractional exclusion statistics
where $n(\Lambda)\sim\frac{1}{g}$ $(~z\rightarrow\infty,\  g={\alpha-\frac{1}{4}}) $. However, for $\alpha\le\frac{1}{4}$
 behavior of $n(\Lambda) $ is similar to a bosonic gas and so we cannot exclude condensed states. This is an interesting possibility for
 this non-Abelian statistics. It should be noted that in this case the non-Abelian statistics has a completely different behavior compared to the
 Abelian fractional exclusion statistics.

 In the thermodynamic limit, the internal energy and particle
number of an exclusion gas in a $D$-dimensional box of volume
$L^D$ with the dispersion relation $\epsilon=ap^{\sigma}$ can be
written as
 \bea
 U=\int_{0}^{\infty}\epsilon \
 n(\Lambda)\Omega(\epsilon)d\epsilon,~~~~~~
 N=\int_{0}^{\infty}n(\Lambda)\Omega(\epsilon)d\epsilon.\label{pn}
 \eea
Neglecting the spin of the particles,
$\Omega(\epsilon)=\frac{A^D}{\Gamma(\frac{D}{2})}\epsilon^{D/\sigma
-1}$ will be  the density of the single particle state for the
system. $A=\frac{L\sqrt{\pi}}{a^{1/\sigma} h}$ is a constant and
 we will set it equal to one $(A=1)$ for simplicity. Obtaining  the
internal energy and the particle number of the ideal non-Abelian
gas for an arbitrary fractional parameter will not be possible
because Eq. (\ref{lambda}) is not solvable with respect to
$\Lambda$ for
 an arbitrary value of fractional parameter. One can change the
integrating variable $\epsilon$ to $\Lambda$ by using the
following relations
 \bea
 \epsilon=\frac{1}{\beta}{\ln}{[\frac{z\sqrt{1+\sqrt{\Lambda}}}{(\Lambda-1)\Lambda^{\alpha-1}}]},~~~~\frac{d}{d\epsilon}\ln\Lambda=-\beta
 n(\Lambda),
 \eea
where $z=\exp(\beta\mu)$ is the fugacity of gas. Thus, the
internal energy and particle number are given by
 \bea
 U=-\frac{\beta^{-(D/\sigma)-1}}{\Gamma(\frac{D}{2})}\int_{\Lambda_{0}}^{1}
 \{\ln[\frac{z\sqrt{1+\sqrt{\Lambda}}}{(\Lambda-1)\Lambda^{\alpha-1}}]\}^{D/\sigma}d(\ln\Lambda),\nonumber\\
 N=-\frac{\beta^{-D/\sigma}}{\Gamma(\frac{D}{2})}\int_{\Lambda_{0}}^{1}
 \{\ln[\frac{z\sqrt{1+\sqrt{\Lambda}}}{(\Lambda-1)\Lambda^{\alpha-1}}]\}^{(D/\sigma)-1}d(\ln\Lambda),\label{UN}
 \eea
where $\Lambda_{0}$ is the zero energy grand partition function
which satisfies
${z\sqrt{1+\sqrt{\Lambda_{0}}}}={(\Lambda_{0}-1)\Lambda_{0}^{\alpha-1}}$
and $\epsilon=\infty$ corresponds to $\Lambda=1$.
\section{Thermodynamic geometry of non-Abelian statistics}
Now, we are going to construct the thermodynamic geometry of an
ideal non-Abelian statistic gas. One can use various
thermodynamic potentials for obtaining the metric elements of the
thermodynamic parameter space. The appropriate representation in
the present case belongs to the second derivatives of the
logarithm of partition function, $\cal{Z}$, with respect to the
thermodynamic parameters $\beta=1/k_{B}T$ and
$\gamma=-\mu/k_{B}T$. Therefore, the metric elements are given by
 \bea\label{M1} g_{ij}=\frac{\partial^{2}\ln
    \cal{Z}}{\partial \beta^{i}\partial\beta^{j}}.
    \eea
The thermodynamic parameter space is a two-dimensional (2D) space
$(\beta^{1},\beta^{2})=(\beta, \gamma)$. For computing the
thermodynamic metric, we select one of the extended variables as
the constant system scale. We will implicitly pick volume in
working with the grand canonical ensemble. The metric elements of
thermodynamic space of an ideal non-Abelian statistic gas are
given by \bea
     \label{vv} &&G_{\beta\beta}=\frac{\partial^{2}\ln\cal{Z}}{\partial
    \beta^{2}}=-(\frac{\partial U}{\partial\beta})_{\gamma}\nonumber\\
   &&=\frac{(\frac{D}{\sigma} +1)\beta^{-[(D/\sigma)+2]}}{\Gamma(\frac{D}{2})}\int_{\Lambda_{0}}^{1}
 \{\ln[\frac{z\sqrt{1+\sqrt{\Lambda}}}{(\Lambda-1)\Lambda^{\alpha-1}}]\}^{D/\sigma}d(\ln\Lambda),\nonumber\\
    &&G_{\beta\gamma}=\frac{\partial^{2}\ln\cal{Z}}{\partial\gamma\partial\beta}=-(\frac{\partial N}{\partial \beta})_{\gamma}\nonumber\\
    &&=\frac{\frac{D}{\sigma}\beta^{-[(D/\sigma)+1]} }{\Gamma(\frac{D}{2})}\int_{\Lambda_{0}}^{1}
 \{\ln[\frac{z\sqrt{1+\sqrt{\Lambda}}}{(\Lambda-1)\Lambda^{\alpha-1}}]\}^{{D\over\sigma}-1}d(\ln\Lambda),\ \ \\
     &&G_{\gamma\gamma}=\frac{\partial^{2}\ln\cal{Z}}{\partial\gamma^{2}}=-(\frac{\partial N}{\partial\gamma})_{\beta}\nonumber\\
     &&=\frac{\beta^{-D/\sigma}}{\Gamma(\frac{D}{2})}\int_{\Lambda_{0}}^{1}
 \frac{1}{n}\frac{\partial
 n}{\partial\Lambda}\frac{\partial\Lambda}{\partial\gamma}\{\ln[\frac{z\sqrt{1+\sqrt{\Lambda}}}
 {(\Lambda-1)\Lambda^{\alpha-1}}]\}^{(D/\sigma)-1}d(\ln\Lambda).\nonumber
    \eea
Now we can evaluate the Christoffel symbol,
$\Gamma_{ijk}=\frac{1}{2}\left(g_{ij,k}+g_{ik,j}-g_{jk,i}\right)$,
using the derivative of the metric elements with respect to the
thermodynamic parameter. Therefore, one can evaluate the
well-known Riemann tensor, Ricci tensor, and finally, the Ricci
scalar, which will be called the thermodynamic curvature because
of the identity of the constructed geometry.

\begin{figure}[t]
    \center
    \includegraphics[width=0.98\columnwidth]{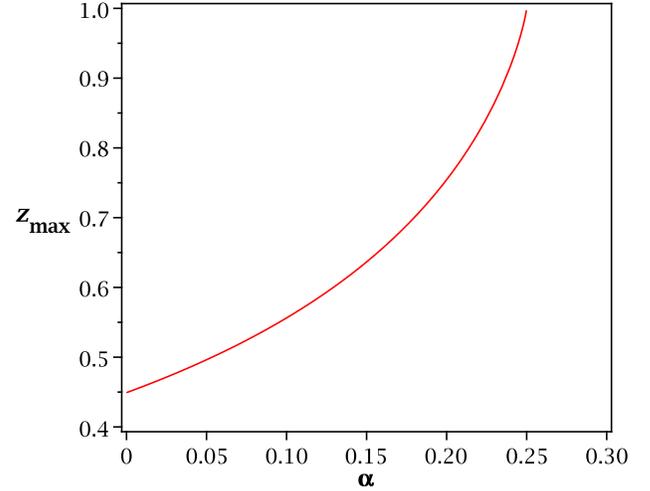}\\
    \caption{(Color online) Maximum value of fugacity of an
    ideal non-Abelian gas as a function of fractional parameter.}\label{figure1}
   \end{figure}

\begin{figure}[t]
    \center
    \includegraphics[width=1.1\columnwidth]{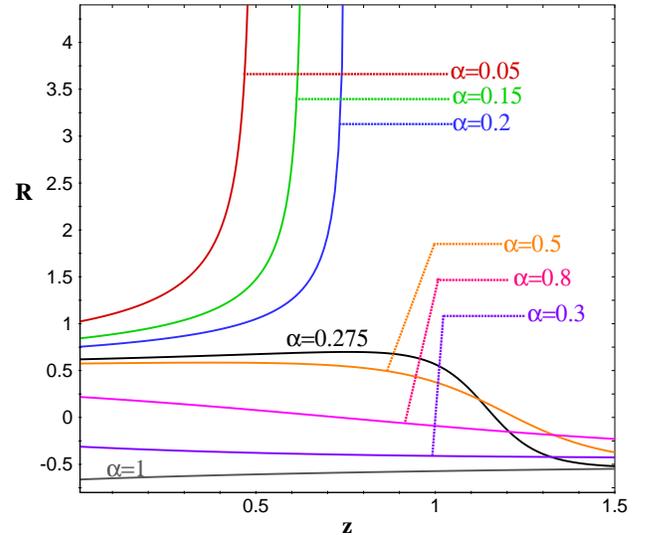}\\
    \caption{(Color online) Thermodynamic curvature of an ideal non-Abelian gas as a
    function of fugacity for isotherm processes $(\beta=1,D/\sigma=3/2)$. $R$ is
    singular at $z=z_{\max}$ for fractional parameter in the interval $0\leq\alpha\leq0.25.$}\label{figure2}
   \end{figure}

Before considering the thermodynamic geometry of the system, we
explain a subtle point about the
fugacity of an ideal non-Abelian statistic gas. We remember that
the fugacity of an ideal boson gas has an upper bound, where the
Bose-Einstein condensation occurs. In a recent paper, we showed
that the fugacity of an ideal gas with particles obeying a special
kind of fractional statistics has an upper bound, too, which is
related to the fractional parameter values \cite{Mirza1}. We have investigated the thermodynamic geometry of
Abelian Haldane fractional exclusion gas in more detail to find
no  upper bound on fugacity and, therefore, no condensation to
occur. Here, we consider the existence of the upper bound on
fugacity of a non-Abelian gas. For zero energy particles, the
right hand side of Eq. (\ref{lambda}) is replaced by the fugacity of
gas. Exploring a condition for the maximum value of fugacity
motivated us to differentiate Eq. (\ref{lambda}) with respect to
$\Lambda$. Thus, $\frac{d z}{d\Lambda}=0$ represents a
 relation  between the fractional parameter and $\Lambda$,
 \bea
 \Lambda=\frac{32\alpha^{2}-40\alpha+9+\sqrt{64\alpha^{2}-80\alpha+17}}{2(4\alpha-1)^{2}},
 \eea
Substituting the above equation in Eq. (\ref{lambda}) reveals that
 the maximum point appears only in the interval $0\leq\alpha\leq 0.25$ as
  depicted in Fig. \ref{figure1}.  For other values of fractional parameter out of this interval, there is
  no upper bound on fugacity.
The thermodynamic geometry of an ideal non-Abelian gas is
presented in Fig. \ref{figure2}. One can observe that the
thermodynamic curvature of a non-Abelian gas has a singularity for
fractional parameters in the interval $0\leq\alpha\leq 0.25$ at
relevant maximum values of fugacity.
It should be mentioned that the statistical interaction of an
ideal non-Abelian gas is attractive in the full physical range
for the  fractional parameter in the specified interval, while for
other values of fractional parameter, the thermodynamic curvature
tends toward negative values at the low temperature limit (quantum
limit) and the statistical interaction becomes repulsive. It seems
that the particles obeying non-Abelian fractional statistics with
the fractional parameter $\alpha>0.25$ at $T=0$ exhibit  Fermi
surfaces such as Haldane fractional exclusion statistics
\cite{Mirza1}.

\section{Condensation of non-Abelian statistic gas}
\begin{figure}[b]
    \center
    \includegraphics[width=1\columnwidth]{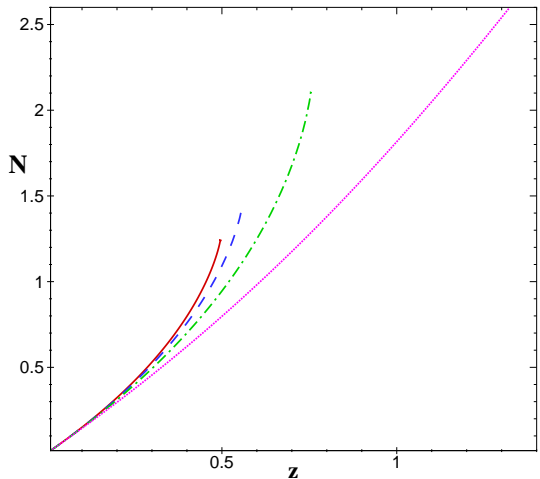}\\
    \caption{(Color online)  Particle number of an ideal
    non-Abelian gas with different fractional parameter (red solid line, $\alpha=0.05$; blue dashed line $\alpha=0.1$,
    green dash-dotted line $\alpha=0.2$; and purple dotted line, $\alpha=0.4$) as a function of fugacity for isotherm processes. $(\beta=1, \  D/\sigma=3/2).$}\label{figure3}
   \end{figure}

\begin{figure}[b]
    \center
    \includegraphics[width=1\columnwidth]{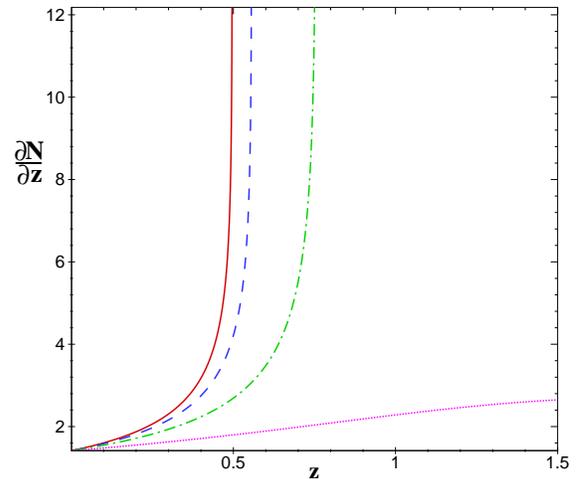}\\
    \caption{(Color online) The derivative of particle number of an ideal
    non-Abelian gas with different fractional parameter (red solid line, $\alpha=0.05$; blue dashed line $\alpha=0.1$;
    green dash-dotted line $\alpha=0.2$; and purple dotted line $\alpha=0.4$) as a function of fugacity for isotherm processes. $(\beta=1, \  D/\sigma=3/2).$}\label{figure4}
   \end{figure}

It is well known that the thermodynamic curvature of an ideal
boson gas is singular at $z=1$. Also, the particle number as a
function of fugacity has an infinite slope at the maximum value of
fugacity where the Bose-Einstein condensation occurs. In the case
of ideal non-Abelian gas, it is evident from Figs. \ref{figure3} and \ref{figure4}
that the particle number for each value of the fractional
parameter in the foregoing interval has an infinite slope and a
finite value at the relevant maximum point of fugacity. In other
words, we can use an analytical method to show that the derivative
of the particle number with respect to fugacity is singular at the
maximum value of fugacity, which indicates that the maximum value
is a critical value of fugacity. According to  Eq. (\ref{pn}) the derivative of
particle number is derived
 \bea
 \frac{\partial N}{\partial z}=\int_{0}^{\infty}\frac{\partial n(\Lambda)}{\partial \Lambda}\frac{\partial\Lambda}{\partial z}\Omega(\epsilon)d\epsilon.
 \eea
 Since $\frac{\partial z}{\partial\Lambda}|_{z_{\max}}=0$, derivative of the  particle number diverges at maximum value of fugacity.
 Therefore, we argue that there
is a condensation for a non-Abelian gas at maximum value of
fugacity for the fractional parameter in a specific interval.
  It has been shown that the phase transition
  temperature for an ideal boson gas has a finite value
  for $\frac{D}{\sigma}>1$. Now we  explore the
  possibility of condensation in various cases for an ideal non-Abelian gas. One
  can obtain the phase transition temperature by using the particle
  number relation  at $z=z_{\max}$ in Eq. (\ref{UN}),
 \bea
 k_{B}T_{c}=\frac{ah^{\sigma}}{\pi^{\sigma/2}}\left(-\frac{\int_{\Lambda_{0}}^{1}\{\ln[\frac{z_{\max}\sqrt{1+\sqrt{\Lambda}}}{(\Lambda-1)
 \Lambda^{\alpha-1}}]\}^{(D/\sigma)-1} d(\ln\Lambda)}{n\Gamma(D/2)}\right)^{-\sigma/D}
 \eea
where $T_{c}$ is the phase transition temperature and
$n=\frac{N}{L^D}$ is the particle density  assumed to be constant.
 \begin{figure}[t]
    \center
    \includegraphics[width=1\columnwidth]{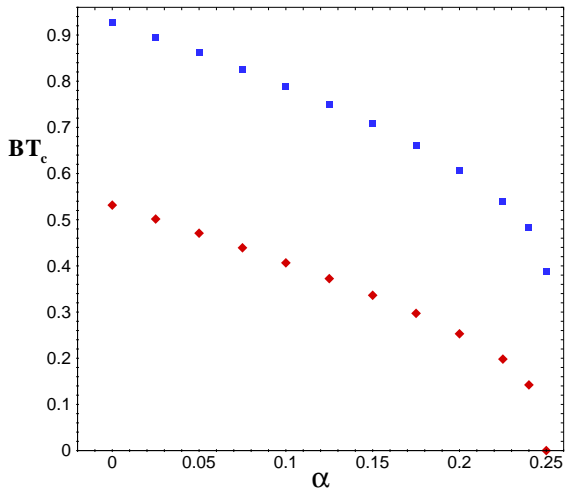}\\
    \caption{(Color online) Phase transition temperature as a function of fractional parameter for non-relativistic particles in $(D=3)$ (blue square
    symbols) and $(D=2)$ (red diamond symbols). $B=\frac{k_{B}\pi}{ah^{2}n^{2/D}}$ is a constant.}\label{figure5}
   \end{figure}
 Figure \ref{figure5} shows that there is a finite
 phase transition temperature in three and two spatial
 dimensions for a nonrelativistic, ideal non-Abelian gas
 with a fractional parameter in the  interval mentioned above. In the two dimensional
 case, for $\alpha=0.25$, the phase transition temperature tends to zero, $T_{c}=0$, while the
  maximum value of fugacity is $z_{\max}=1$. This point is similar
  to a boson gas which has no condensation in two dimensions for
  nonrelativistic particles. For ultrarelativistic particles, one can find a finite
  critical temperature in all dimensions except in $D=1$ for $\alpha \geqslant 0.25$. It is an interesting
  phenomenon that, unlike the case of the boson gas, the condensation for a non-Abelian gas  can
  occur
  in even two spatial dimensions with the  particles in a nonrelativistic regime but in all  dimensions in an ultrarelativistic regime.

The interval of fractional parameter in which there is an upper
bound on fugacity may be related to the selection of $M$ types of
particles and $k$ types of pseudoparticles. We selected the
simplest form of non-Abelian statistics. For the other values of
$M$ and $k$, we encountered a certain degree of  complexity when
trying to  obtain  the distribution function and the ensuing
evaluations. To answer the initial question of this paper,
selection of the simplest case led us to the non-Abelian exclusion
statistics which could  be condensate  in even  low dimensions.
\section{conclusion and discussion}
We considered the non-Abelian statistics, which has $k=2$. We notice that if one sets $G_{iA}=0$, which is the non-Abelian part of  the statistics, we will have
\bea
(\Lambda-1)\Lambda^{\alpha}=z.
\eea
The above equation describes the non-Abelian statistics, where $\alpha$ is a parameter and takes values $0\le\alpha\le 1$. As we know, the Moore-Read state in addition to $k=2$ has a free parameter $M=q-1$, which set the filing fraction, $\nu=k/(M k+2)=1/(M+1)$,  and the related fractional parameter is $\alpha=[(k-1)M+2]/2(k M+2)=(M+2)/4(M+1)$, which is bounded from below by $1/4$ \cite{sathya}. Therefore, Eq. (\ref{lambda}) for $\alpha>1/4$ describes the Moore-Read states and according to the behavior of thermodynamic curvature and mean occupation number in the low temperature limit, the statistics is fractional exclusion such as Haldane statistics, and therefore there is no condensation for $\alpha>1/4$.  We have also explored  $0\le \alpha \le 1/4$ where the  theory is still well defined and consistent, however,  the non-Abelian statistics is boson like and has no exclusion property.  The statistical interaction is attractive in full physical range and it has singularity for specified values of fugacity where the condensation occurs. It should be noted that the non-Abelian character of the gas for $0\leq \alpha \leq 1/4$ should be investigated more carefully in the future and we cannot rely on the results from fractional exclusion statistics.

Thermodynamic geometry of an ideal non-Abelian statistic gas worked out and one can observe that
for $\alpha\le1/4$, there is an upper bound for fugacity of gas, where the thermodynamic curvature is
singular. Also, the particle number has a finite value and infinite slope at $z=z_{\max}$. Therefore, we argued that there is a
phase transition such as Bose-Einstein condensation for specified values of fractional parameter. Phase transition temperature was
derived and it was shown that there is a finite temperature, even for low dimensions.

Whereas the non-Abelian  statistics has many interesting
relationships  with the fractional quantum Hall effect and the
non-Abelians play a major role in the $q$-pfaffian Hall states
\cite{Moore} and also due to their candidacy for
constructing the topological quantum computers \cite{kitaev}, the
condensation of these kinds of non-Abelian statistics in $3D$ and
$2D$ can be an outstanding phenomenon in this field and may point
to new states of matter.

\section*{Acknowledgment}
We would like to acknowledge the referees for their valuable comments, which brought more clarity to the paper.

%%%%%%%%%%%%%%%%%%%%%%%%%%%%%%%%%%%%%%%%%%%%%%%%%%%%%%%%%%%%%%%%%%%%%%%%%%%%

\end{document}